\documentstyle{elsart}
\input psbox.tex
\begin{document}
\begin{frontmatter}

\title{ON THE THIRD LAW OF BLACK \\
 HOLE DYNAMICS}
\author{Naresh Dadhich \thanksref{emal}}
\thanks[emal]{E-mail : naresh@iucaa.ernet.in; Fax : 91-212-350760} 
\address{Inter-University Centre for Astronomy \& Astrophysics,
 Post Bag 4, Ganeshkhind, Pune - 411 007.}
\author{ K. Narayan \thanksref{emal1}}
\thanks[emal1]{E-mail : narayan@niharika.phy.iitb.ernet.in}
\address{ Department of Physics, IIT Powai, Bombay - 400 076}

\begin{abstract}
\noindent The  third  law of black hole dynamics states that  the  surface 
gravity (temperature) of black hole cannot be reduced to zero  in 
finite sequence of physical interactions. We argue that
the same is true when surface gravity is replaced by gravitational 
charge. 
We demonstrate that the  prescribed  window for infalling energy and  radiation 
pinches off as extremality ($M^2 = a^2 + Q^2 $) is approached. 
\end{abstract}
\end{frontmatter}

\vfill
\begin{flushright} IUCAA-31/97 \end{flushright}
\vfill
\newpage

\noindent  Parallel  to the laws of thermodynamics, the laws of black  hole 
dynamics (BHD) were formulated comprehensively by Bardeen, Carter  
and  Hawking \cite{Bard73}. The identification of temperature with  surface 
gravity  of the hole was clinched by  the Hawking 
radiation  \cite{Hawk74}  
which  followed  from application of quantum  theory  in  general 
relativity. Subsequently Israel has given the precise formulation 
and proof of the third law \cite{Isra86}. Although the law has been  proved 
in  an elegant way using sophisticated global analysis,  a  clear 
and  straight  forward demonstration of its working has  its  own 
merit and usefulness \cite{Wald74}. We shall argue that the law can also
be stated by replacing surface gravity by  gravitational
charge enclosed by the horizon. Gravitational charge of
a hole can be defined by the flux of red-shifted proper
acceleration across the closed 2-surface defined by the
horizon \cite{Dadh89,Chel90} and value of the acceleration at the
horizon defines the surface gravity. Thus the two
quantities are intimately related. By considering variation in gravitational
charge and applying the area non-decrease theorem,  we 
would  like to exhibit how it cannot 
be  reduced to zero in finite sequence of physical processes.  It 
turns  out  that as extremality ($M^2 = a^2 + Q^2 $)  is  approached,  the 
window  for allowed range of parameters of infalling  energy  and 
radiation pinches off. \\

\noindent The  third law of thermodynamics has two essentially  equivalent 
statements;  (a) isothermal reversible processes turn  isentropic 
as  temperature of a system approaches zero, and (b)  temperature 
cannot be reduced to zero in finite number of physical operations 
\cite{Wils57}.  There is yet another stronger version due to  Planck  which 
states  that  the  entropy of any system  tends  to  an  absolute 
constant,  which  can be taken as zero, as temperature  tends  to 
zero. 

\noindent In  the  case  of a black hole, surface gravity  tends  to  zero 
either as $M^2 \longrightarrow a^2 + Q^2$
for charged and rotating black hole   or $M \longrightarrow \infty$,          
where $M, a = J/M $, and $Q$ refer respectively  to  mass,  specific 
angular momentum and electric charge of the hole. In either  case 
area  of  the hole, which is analogus to entorpy, does not  go  to 
zero.  Hence Planck's  version has no analogue in classical  BHD. 
However it has been well recognised that thermodynamic description will
not be tenable for extremal black hole. In particular it   has  recently  been  argued  on  topological   and   quantum 
considerations  \cite{Hawk95,Teit95}  
that extremal ($M^2 = a^2 + Q^2$) black  hole  is 
qualitatively   different  from  non-extremal  black  hole.   The 
identification  of area with entropy is true only for the  latter 
and is not true for the former. For the extremal case entropy can 
be  deduced separately and it vanishes, thus according well  with 
Planck's  version.  The  conclusion is that  area  of  the  event 
horizon does not always measure  entropy of the hole. It  does 
so only for non-extremal black holes while for extremal case area 
is  though finite and non-zero but entropy turns out to be  zero. 
This   is  a  new  proposal  based  primarily  on   non-classical 
considerations. 

\noindent  Let us look at the  familiar 
mass formula for a black hole, 
\begin{equation}
 M = \frac{\kappa}{4 \pi} A + 2wJ + Q \phi 
\end{equation}

\noindent where  all  symbols have the usual  meaning.  In  thermodynamical 
sense  the  first term on the right should measure  the  internal 
energy  (enthalpy) while the remaining two terms  indicate  the 
work done on the hole. The internal energy, $M_I$, can be identified 
with  the  effective  gravitational  charge  of 
 the  hole \cite{Dadh89}, which 
is defined by \cite{Chel90}, 

\begin{equation}
M_g = \frac{1}{4 \pi} \int {\bf g. d s} =
M - \frac{a^2}{r_+} - \frac{Q^2}{r_+} = (M^2 - a^2 - Q^2)^{1/2} \label{eq:gds}
\end{equation} 

\noindent where the integration is taken over the closed 2-surface defined by
the  horizon,  and  ${\bf g} = - \alpha {\bf \bigtriangledown} (ln \alpha), ~ \alpha$   is  the  norm  of  the 
timelike   corotating   vector ${\partial \over \partial t} +
w {\partial \over \partial \varphi} $.  This   follows   from 
application   of   the  Gauss  theorem  to   red-shifted   proper 
acceleration ${\bf g}$, its norm represents surface gravity when
evaluated at the horizon. This is in fact the Komar integral \cite{Kom59} for
the corotating timelike vector ${\partial \over \partial t} + w {\partial
\over \partial \varphi}$, over the horizon. It is therefore the Komar
mass of the hole evaluated at the horizon \cite{Chel90}. Unfortunately this vector
is not Killing in general and hence the integral does not yield an invariant
mass. However it does give a good measure for $r \longrightarrow r_+$
and $r \longrightarrow \infty$, because in these limits the vector
does tend to be Killing. It will give 

\begin{equation}
 M_g = M_I = \frac{\kappa}{4 \pi} A = \frac{2 M^2_{ir}}{M} - M \label{eq:eul}
\end{equation}

\noindent where 

\begin{equation}
M^2_{ir} = \frac{1}{2} M r_+ = \frac{A}{16 \pi},~~
r_+ = M + (M^2 - a^2 - Q^2)^{1/2}. 
\end{equation} 

\noindent $M_{ir}$ is called the irreducible mass 
of the hole \cite{Misn73}.
Note that $M_I = (\kappa/4 \pi) A$
is not a defining relation for $M_g$, which is the measure
of flux of ${\bf g}$ across the closed 2-surface defined
by the horizon. However ${\bf g}$ also defines the surface
gravity and we have $M_g = M_I$.

\noindent $M_g$  tends  to zero as extremality is reached. That is gravitational charge contained inside the horizon vanishes
for an extremal black hole. If entropy of the hole is to
depend upon gravitational charge contained in the hole, it should also
vanish for an extremal hole as argued in \cite{Hawk95,Teit95}.
There is however no relation connecting entropy with gravitational
charge. All we can say 
is  that  area  of the horizon is not a  measure  of  entropy  of 
extremal hole.  

\noindent Here  an analogy can be drawn between extremal and  non-extremal  
black holes, and photons and ordinary particles, indicating  their 
characteristic difference. The analogy suggests that gravitational
charge  $M_g$
corresponds  to rest mass, we have $M^2_g = M^2 - a^2 - Q^2 $ analogus
$m^2 = E^2 - p^2$. For photon rest mass energy vanishes and  its  entire 
energy is kinetic while for extremal hole its gravitational
charge vanishes and its entire energy is  
-  rotational and/or
electromagnetic. The third law simply states that a non-
extremal black hole cannot be converted into an extremal one ( an 
ordinary  particle  cannot be accelerated into a photon).  Like 
photon  it has to be born like that. For photon the  converse  is 
also  true,  i.e a photon cannot be converted  into  an  ordinary 
particle. This does not appear to be the case for black hole, for 
nothing prohibits classically to add mass to an extremal hole  to 
make  it  non-extremal.  It has however been  argued  on the basis
of radiation properties of black holes 
that  extremal hole can be  in  equilibrium  with 
thermal radiation at any temperature and hence it can radiate  at 
any rate independent of temperature. It can thus be thought  that 
extremal black hole always radiates in such a way when matter and 
radiation fall into it so as to keep itself extremal \cite{Hawk95}. This  
seems to make the analogy  with  photon  perfect. 

\noindent Another  equivalent  statement  of the third law  could  be  that 
gravitational  charge  of the hole cannot be  reduced 
to  zero  in finite sequence of physical  processes.  Further  as 
argued by Hawking, et.al \cite{Hawk95}, that if it happens to be zero, then 
no finite sequence of physical operations can make it non-zero. 
The latter statement would however have to be formulated as a 
separate law. That  means  black  holes are  characterised  
into  two  distinct 
classes  by  vanishing and non-vanishing of  their  gravitational 
charge  and the two classes are qualitatively different and  non-
interchangable  exactly in the similar sense as photons are  from 
ordinary particles. 

\noindent From eqn. (\ref{eq:eul}) let us consider variation in gravitational charge, 

\begin{equation}
\delta M_g = - \left(1 + \frac{2M^2_{ir}}{M^2}\right) \delta M
+ 4 \frac{M_{ir}}{M} \delta M_{ir}. 
\end{equation}

\noindent This will tend to zero as extremality ($M^2 = a^2 + Q^2$)
is approached, both terms tend to the same limit and cancel
each other. The 
process  tends  to  be isenthalpic.  All 
particles that tend to decrease $M_g$ are scattered off by the 
hole as $M^2 \longrightarrow a^2 + Q^2 $. 

\noindent For  simplicity let us now specialise to a rotating hole,  while 
all  our results will hold good even when $Q \not= 0$. If  we  consider           
$\delta M_g \leq 0$ and $\delta A \geq 0$, 
the former implies an upper limit while the  latter  a 
lower limit on $\delta M/\delta J$ falling into the hole,
and then we obtain  the following window for permissible range, 

\begin{equation}
\frac{a}{2Mr_+} \leq \frac{dM}{dJ} \leq \frac{a}{M^2 + a^2} < \infty. \label{eq:inf}
\end{equation} 

\noindent Now  both  lower  and upper limits tend to $1/2M $ as 
$M^2 \longrightarrow a^2$, thus 
completely  pinching  off the window. This  clearly  demonstrates 
that $M_g = M_I$  can never be reduced to zero and all interactions
turn isenthalpic and isentropic (reversible) as extremality is
approached. 

\noindent We shall next consider variation in surface gravity.
The analogue of (\ref{eq:inf}) will now be  

\begin{equation}
\frac{a}{2Mr_+} \leq \frac{dM}{dJ} \leq \frac{aM}
{-M^3 + 3M a^2 - (M^2 - a^2)^{3/2}} < \infty \label{eq:inf1}
\end{equation} 

\noindent where we have used 

\begin{equation}
A = 8 \pi M r_+~,  
\end{equation} 

\begin{equation}
\delta A = 8 \pi \left[\frac{2Mr_+ \delta M - a \delta J}
{(M^2 - a^2)^{1/2}} \right] 
\end{equation} 

\noindent and 

\begin{equation}
\kappa = \frac{(M^2-a^2)^{1/2}}{2Mr_+} = \frac{M_g}{2M r_+} =
\frac{M_g}{4M^2_{ir}} \label{eq:mr}
\end{equation} 

\begin{equation}
\delta \kappa = \frac{\delta M [-M^2 + 3a^2 - \frac{1}{M}
(M^2 - a^2)^{3/2}] - a \delta J}{2M r^2_+ \sqrt{M^2 - a^2}}.
\end{equation}

\noindent Here again either side in (\ref{eq:inf1}) will 
tend to the same limit $1/2M$, 
yielding  the  same  conclusion that surface  gravity  cannot  be 
reduced   to  zero  and  all  interactions  turn  isothermal
and isentropic (reversible) as extremality is approached. 

\noindent Thus  all  interactions  with black  hole  that  point  towards 
extremality   turn  isoenthalpic,  isentropic,   isothermal   and 
reversible  as extremality is approached. No finite  sequence  of 
physical   interactions  can  reduce  the  surface  gravity   and 
gravitational  charge of  black hole to zero.  A  non-extremal 
hole  cannot evole into an extremal one. The converse  of  this 
statement has also been discussed and 
justified by Hawking, et.al \cite{Hawk95} but it  cannot 
be established from these purely classical considerations. 
The arguement 
crucially  rests  on radiation properties of the hole  which  are 
governed by quantum considerations. 

\noindent As demonstrative simple examples, let us consider one, evolution 
of surface gravity of a hole along the isentropic (constant area  
- analogus to adiabatic process in thermodynamics) path as shown 
in Fig.1, and second, along the constant angular momentum path as 
shown in Fig.2. In the former case there is a monotonic evolution 
and which is reversible, while Fig.2 depicts the evolution   from  
extremality  to  decreasing  $a/M$. In the latter case,
$\kappa$ will   increase 
initially as matter is added into a (nearly) extremal hole
and attain maximum value at $a^2/M^2 = \sqrt 3 (2
- \sqrt 3) \approx 0.46 ~ (a/M \approx 0.68)$, and will then 
start  decreasing as mass begins to dominate over rotation.  Here 
the process is irreversible and hence it will not trace the  same 
curve in the reverse direction. 

\noindent There have been several attempts to define quasi-local energy
of black hole spacetimes (for example, \cite{Pen69,Bro93}). 
It is supposed to be a
measure of energy contained inside some compact surface. It may be noted
that gravitational charge as defined in (\ref{eq:gds}) does not agree with the 
quasi-local energy. This point has been specifically discussed in \cite{Chel90}.
In our discussion, it is gravitational charge which acts as a source for
gravitational attraction - the surface gravity. In this respect it
measures a physically meaningful property of the hole. We have demonstrated 
explicitly that it cannot be reduced to zero through any finite sequence
of operations. 

\ack
KN thanks Jawaharlal Nehru Centre
for Advanced Scientific Research, Bangalore for a summer
fellowship to work at IUCAA, Pune. We also thank the referee for
constructive criticism.

\newpage

\begin{figure}
\label{fig1}
\psbox{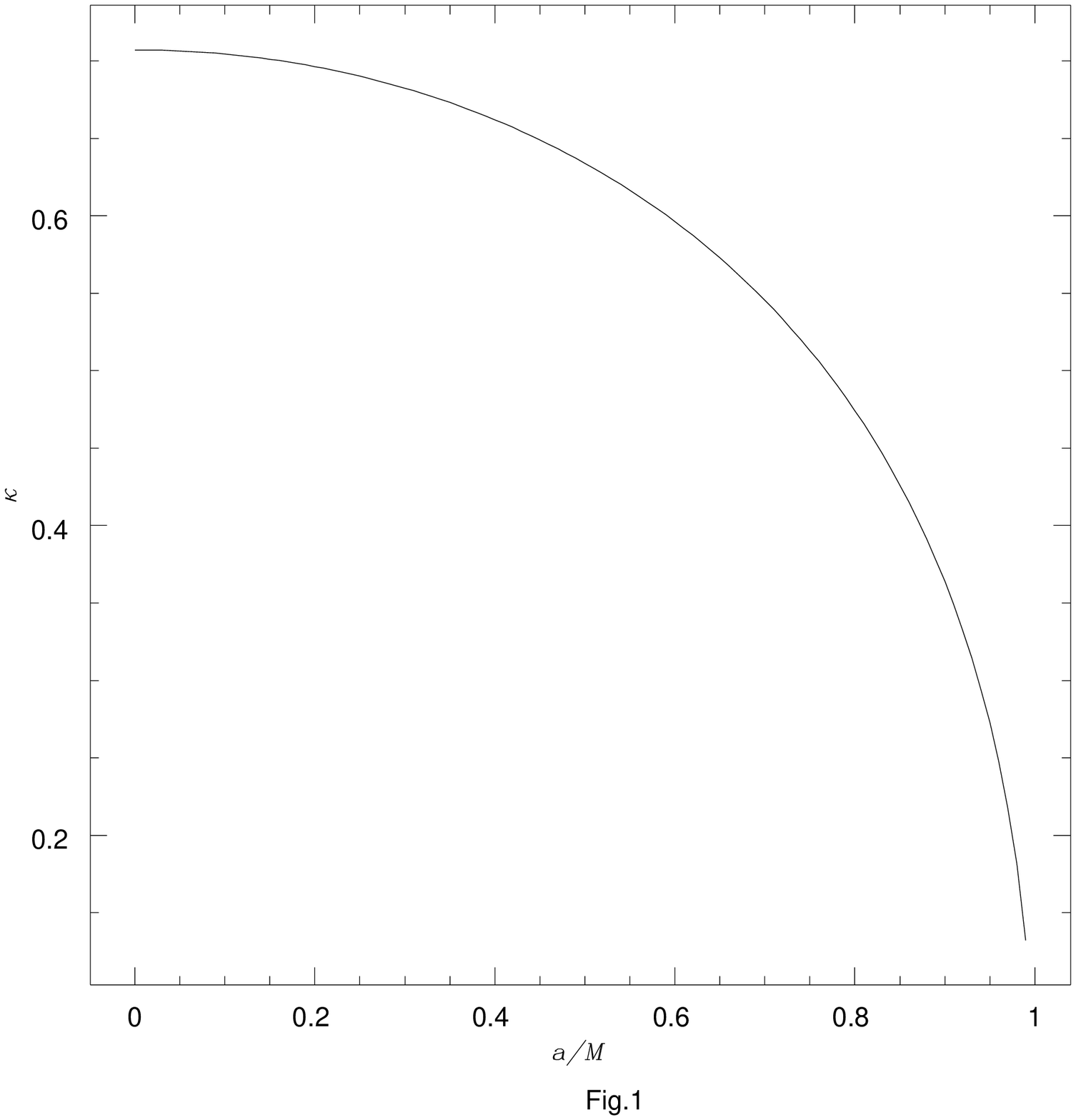}
\end{figure}

\newpage
\begin{figure}
\label{fig2}
\psbox{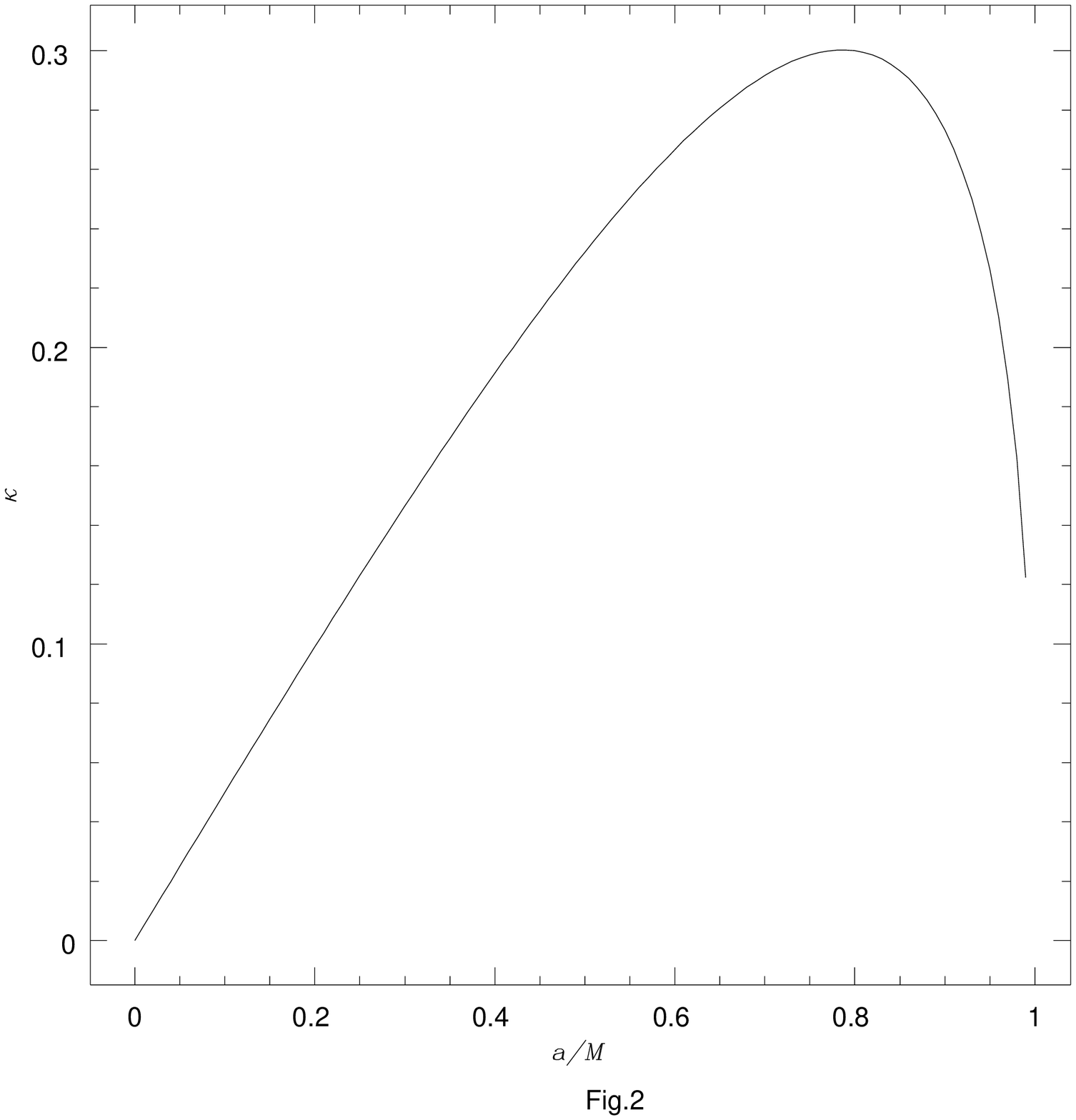}
\end{figure}


\begin{thebibliography}{99}
\bibitem{Bard73} J M Bardeen, B Carter and S W Hawking (1973) Commun.
Math. Phys. {\bf 31}, 161.

\bibitem{Hawk74} S W Hawking (1974) Nature {\bf 248}, 30.

\bibitem{Isra86} W Israel (1986) Phys. Rev. Lett. {\bf 57}, 397.

\bibitem{Wald74} R M Wald (1974) Ann. Phys. {\bf 83}, 548.

\bibitem{Dadh89} N Dadhich (1989) GR-12 Abstracts, p.60.

\bibitem{Chel90} V Chellathurai and N Dadhich (1990) Class. Quantum
Grav. {\bf 7}, 361.

\bibitem{Wils57} A H Wilson, {\it Thermodynamics and Statistical 
Mechanics} (Cambridge University Press, Cambridge, 1957).

\bibitem{Hawk95} S W Hawking, G T Horowitz and S F Ross (1995) Phys.
Rev. {\bf D 51}, 4302.

\bibitem{Teit95} C Teitelboim (1995) Phys. Rev. {\bf D 51}, 4315.

\bibitem{Kom59} A Komar (1959) Phys. Rev. {\bf 113}, 934.

\bibitem{Misn73} C W Misner, K S Thorne and J A Wheeler, {\it Gravitation}
(W H Freeman and Company, San Francisco, 1973).

\bibitem{Pen69} R Penrose (1969) Proc. Roy. Soc. London {\bf A 381}, 53.

\bibitem{Bro93} J D Brown and J W York Jr. (1993) Phys. Rev. {\bf D 47}, 1407.

\end{thebibliography}
\end{document}